\documentclass[12pt,epsfig]{article}
\usepackage{epsfig}
\begin{document} 
 
\def\a{\alpha}
\def\b{\beta}
\def\c{\varepsilon}
\def\d{\delta}
\def\e{\epsilon}
\def\f{\phi}
\def\g{\gamma}
\def\h{\theta}
\def\k{\kappa}
\def\l{\lambda}
\def\m{\mu}
\def\n{\nu}
\def\p{\psi}
\def\q{\partial}
\def\r{\rho}
\def\s{\sigma}
\def\t{\tau}
\def\u{\upsilon}
\def\v{\varphi}
\def\w{\omega}
\def\x{\xi}
\def\y{\eta}
\def\z{\zeta}
\def\D{\Delta}
\def\G{\Gamma}
\def\H{\Theta}
\def\L{\Lambda}
\def\F{\Phi}
\def\P{\Psi}
\def\S{\Sigma}

\def\o{\over}
\def\beq{\begin{eqnarray}}
\def\eeq{\end{eqnarray}}
\newcommand{\gsim}{ \mathop{}_{\textstyle \sim}^{\textstyle >} }
\newcommand{\lsim}{ \mathop{}_{\textstyle \sim}^{\textstyle <} }
\newcommand{\vev}[1]{ \left\langle {#1} \right\rangle }
\newcommand{\bra}[1]{ \langle {#1} | }
\newcommand{\ket}[1]{ | {#1} \rangle }
\newcommand{\EV}{ {\rm eV} }
\newcommand{\KEV}{ {\rm keV} }
\newcommand{\MEV}{ {\rm MeV} }
\newcommand{\GEV}{ {\rm GeV} }
\newcommand{\TEV}{ {\rm TeV} }
\def\diag{\mathop{\rm diag}\nolimits}
\def\Spin{\mathop{\rm Spin}}
\def\SO{\mathop{\rm SO}}
\def\O{\mathop{\rm O}}
\def\SU{\mathop{\rm SU}}
\def\U{\mathop{\rm U}}
\def\Sp{\mathop{\rm Sp}}
\def\SL{\mathop{\rm SL}}
\def\tr{\mathop{\rm tr}}

\def\IJMP{Int.~J.~Mod.~Phys. }
\def\MPL{Mod.~Phys.~Lett. }
\def\NP{Nucl.~Phys. }
\def\PL{Phys.~Lett. }
\def\PR{Phys.~Rev. }
\def\PRL{Phys.~Rev.~Lett. }
\def\PTP{Prog.~Theor.~Phys. }
\def\ZP{Z.~Phys. }


\baselineskip 0.7cm

\begin{titlepage}

\begin{flushright}
UT-07-15
\end{flushright}

\vskip 1.35cm
\begin{center}
{\large \bf
    Determining the Mass for a Light Gravitino
}
\vskip 1.2cm
K. Hamaguchi, S. Shirai and T. T. Yanagida
\vskip 0.4cm

{\it  Department of Physics, University of Tokyo,\\
     Tokyo 113-0033, Japan}

\vskip 1.5cm

\abstract{Gauge mediated supersymmetry breaking scenarios with an ultra-light gravitino of mass $m_{3/2}=1$--10~eV 
are very interesting, since there is no cosmological gravitino problem. We propose a new experimental 
determination of the gravitino mass for such an ultra-light gravitino, 
by measuring a branching ratio of two decay modes of sleptons.}
\end{center}
\end{titlepage}

\setcounter{page}{2}

\section{Introduction}

Gauge mediation of supersymmetry (SUSY) breaking in a hidden to the SUSY standard model (SSM) sector~\cite{Giudice:1998bp}
is very attractive, since it provides us with a natural solution to 
the problems of flavor-changing neutral current and CP-violation in the SSM. 
The gauge mediation predicts a light gravitino ($\tilde{G}_{3/2}$) of mass $m_{3/2}=1~{\rm eV}-10$ GeV,
which is most likely the lightest SUSY particle (LSP).
In particular, an ultra-light gravitino of mass in a range of $1-10$ eV is very interesting, 
since there is no cosmological gravitino problem~\cite{Pagels:1981ke+Viel:2005qj}. 
For such an ultra-light gravitino, the determination of  
its mass will be of crucial importance at the LHC and ILC,
since the small gravitino mass would predict a relatively low SUSY-breaking scale 
$\sqrt{F} = \sqrt{\sqrt{3}m_{3/2}M_P} \simeq 60-200$ TeV,
where $M_P\simeq 2.4\times 10^{18}$ GeV is the reduced Planck scale.
If it is the case, 
it might not be impossible to access the physics for the SUSY 
breaking in future experiments.

Nevertheless, it is very difficult to determine the gravitino mass for the light gravitino. 
The gravitino mass may be determined by measuring the lifetime of the next lightest SUSY particle (NLSP),
since the NLSP lifetime is proportional to $F^2\propto m_{3/2}^2$~\cite{NLSPlifetime,Kawagoe:2003jv}.\footnote{If the gravitino mass is large enough, a measurement of the gravitino mass by kinematical reconstruction may also be possible,
 which serves as a test of supergravity~\cite{SUGRAtest}. However, this is also impossible for the ultra-light gravitino.} For the ultra-light gravitino, 
however, the NLSP lifetime is very short and it will be very challenging  to measure its decay length:
For a slepton NLSP, the decay length is given by $c \tau \simeq 0.55~\mu\mathrm{m}
 (m_{3/2}/1~\mathrm{eV})^2 (m_{\rm NLSP}/200~\mathrm{GeV})^{-5}$, 
whereas the experimental resolution of the impact parameter is 
at least $\mathcal{O}(10)~\mu\mathrm{m}$ or larger \cite{LHC-TDR,TESLA-TDR}.

In this letter we propose a possible independent experimental determination of
the mass for such a light gravitino ${\tilde G}_{3/2}$. 
We consider the case where
the NLSP is the lighter stau ($\tilde{\tau}$), and the 
right-handed smuon ($\tilde{\mu}$) and selectron 
($\tilde{e}$) are slightly heavier than the stau ($\tilde{\tau}$) but lighter than the lightest 
neutralino $\chi^0_1$, $m_{\tilde{\tau}} < m_{\tilde{\mu}/\tilde{e}} < m_{\chi^0_1}$, which is a typical mass spectrum
 in the gauge mediated SUSY breaking scenario (GMSB) as shown in the next section. The smuon/selectron 
then decays into two channels;
one is ${\tilde \ell}\rightarrow {\tilde \tau} +\tau +\ell$ and the other is ${\tilde \ell}
\rightarrow \ell + {\tilde G}_{3/2}$, where $\tilde\ell$ ($\ell$) denotes 
$\tilde{\mu}$ or $\tilde{e}$ ($\mu$ or $e$).
The former process is fixed by the SSM interactions and SUSY-particle masses and the latter
is fixed by the gravitino mass and the smuon/selectron mass.
Therefore, we can determine the gravitino mass
by measuring the ratio of these decay processes, provided that 
the masses of relevant SUSY particles are known.

\section{Gauge-mediation model}

We consider a simple gauge-mediation model, where
a SUSY breaking field $S$ couples to $N$ pairs of
messenger chiral superfields, $\psi$ and $\bar{\psi}$, 
which transform as
${\bf 5}$ and ${\bf 5}^*$ under the $SU(5)_{\rm GUT}$:
$W = k\psi\bar{\psi}S$. The $S$ field develops a vacuum expectation value $k\langle S \rangle = M + \theta^2 F$, where $M$ is the messenger mass. 
We assume an ultra-light gravitino of mass $1-10$ eV, and therefore it is natural to assume a direct gauge-mediation, that is
$F/k=\sqrt{3}m_{3/2}M_P$. We choose $k=1$ in the following discussion, for simplicity.

With the above conditions the low-energy spectrum of the SUSY particles including
the gravitino mass are determined by 5 parameters, $F$, $M$, $N$, 
$\tan\beta$, and ${\rm sgn}(\mu) = \pm 1$. We consider $N\le 4$ to maintain the perturbative
unification of the standard model (SM) gauge couplings.

Let us now discuss the spectrum of the SSM SUSY particles. In the GMSB models gaugino masses are generated 
from loop diagrams of the messengers~\cite{Giudice:1998bp}. At the one-loop level, gaugino masses are given by
\begin{equation}
M_{a} = \frac{N\alpha_a}{4\pi}\Lambda g(x)~ ~(a=1,2,3),\label{eq:gaugino_mass}
\end{equation}
where $\Lambda = F/M$, $x=F/M^2$, $\alpha_1=5 \alpha _{\rm EM}/(3 \cos^2\theta_{W})$ and 
\begin{equation}
g(x) = \frac{1}{x^2}[(1+x)\log(1+x)+(1-x)\log(1-x)].
\end{equation}
Scalar masses, at the two loop level, are written as
\begin{equation}
m^2_{\phi_i}=2N\Lambda ^2 \sum_a \left(\frac{\alpha_a}{4\pi}\right)^2 C_a (i) f(x), \label{eq:scalar_mass}
\end{equation}
where $C_a(i)$ are Casimir invariants for the particle $\phi_i$ 
($C_1(i) = 3Y_i^2/5$)
and
\begin{equation}
f(x) = \frac{1+x}{x^2}\left[ \log(1+x) - 2 {\rm Li}_2(x/[1+x]) + \frac{1}{2} {\rm Li}_2(2x / [1+x]) \right] + (x \rightarrow -x).
\end{equation}
For $x<1$, both $f(x)$ and $g(x)$ are ${\cal O}(1)$. Detailed behaviors of these functions are given in~\cite{Giudice:1998bp}.
$F < M^2$ should be satisfied, since otherwise messenger scalars are tachyonic and the SM gauge symmetry is spontaneously broken. We consider the case of $F < 0.8 M^2$ in the present analysis. The masses in Eq.(\ref{eq:gaugino_mass}) and Eq.(\ref{eq:scalar_mass}) are given at the messenger scale. To calculate the decay rate
of the scalar leptons, we need to evaluate the on-shell masses and mixings for the SUSY particles. We 
should evolve the renormalization group equations down to the electroweak scale and calculate the neutralino mixing and scalar's left-right mixing.
To evaluate those values, we have used the program 
{\verb SOFTSUSY } 2.0.11~\cite{Allanach:2001kg}. In our computation, we have chosen ${\rm sgn}(\mu) = +1$.

\begin{figure}[t]
\hspace{-3mm}
\begin{tabular}{cc}
\begin{minipage}{0.5\hsize}
\begin{center}
\rotatebox{-90}{\epsfig{file=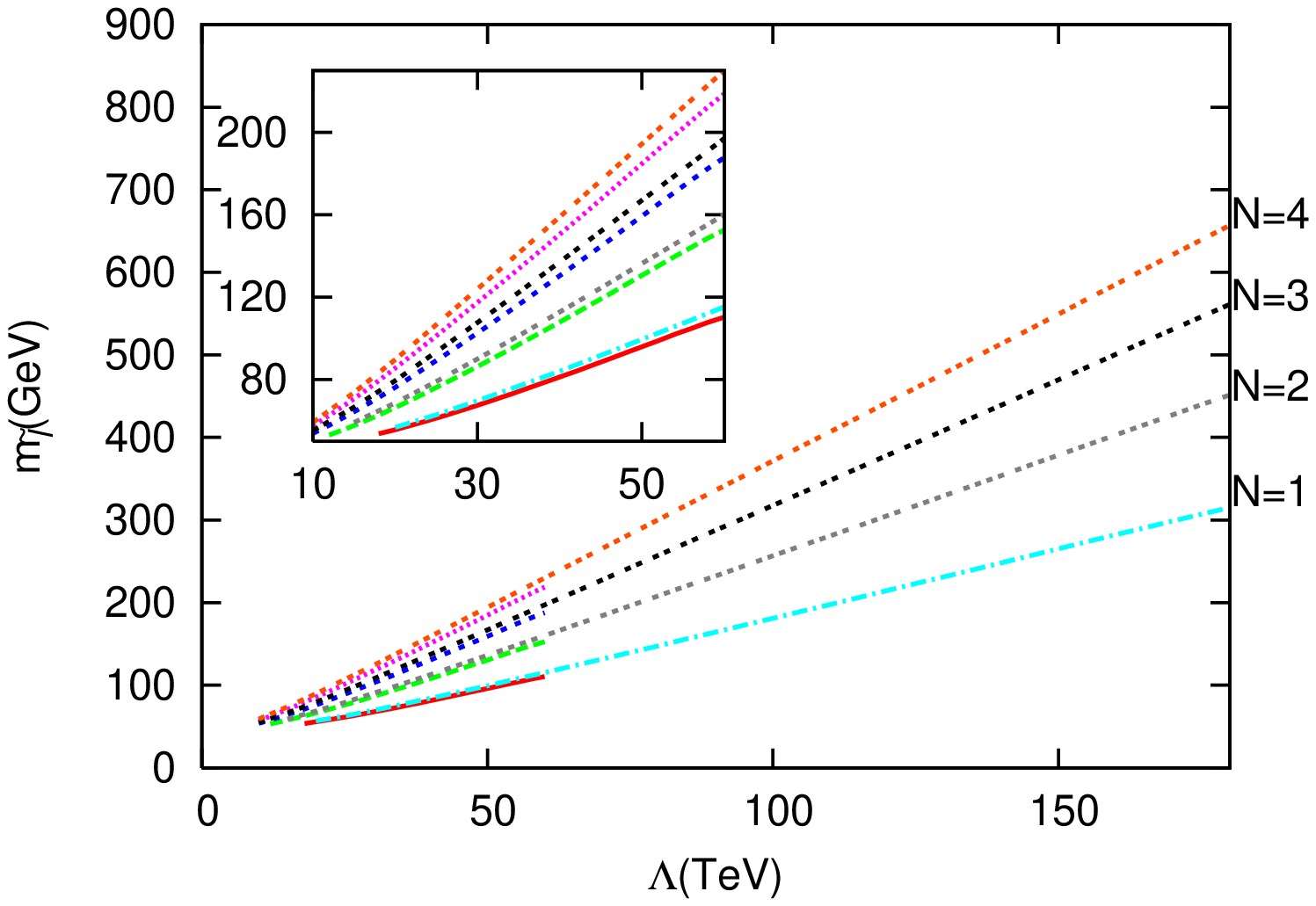,bb=251 55 554 487, clip,width=5.5cm}}
\caption[]{$m_{\tilde{\ell}}$'s dependence on $\Lambda$ for $\tan \beta=10$. The lines with a short range are for $m_{3/2} = 1~ {\rm eV}$, and the others for $m_{3/2} = 10~ {\rm eV}$.
 }
\label{fig:slepton}
\end{center}
\end{minipage}
\hspace{5mm}
\begin{minipage}{0.5\hsize}
\begin{center}
\rotatebox{-90}{\epsfig{file=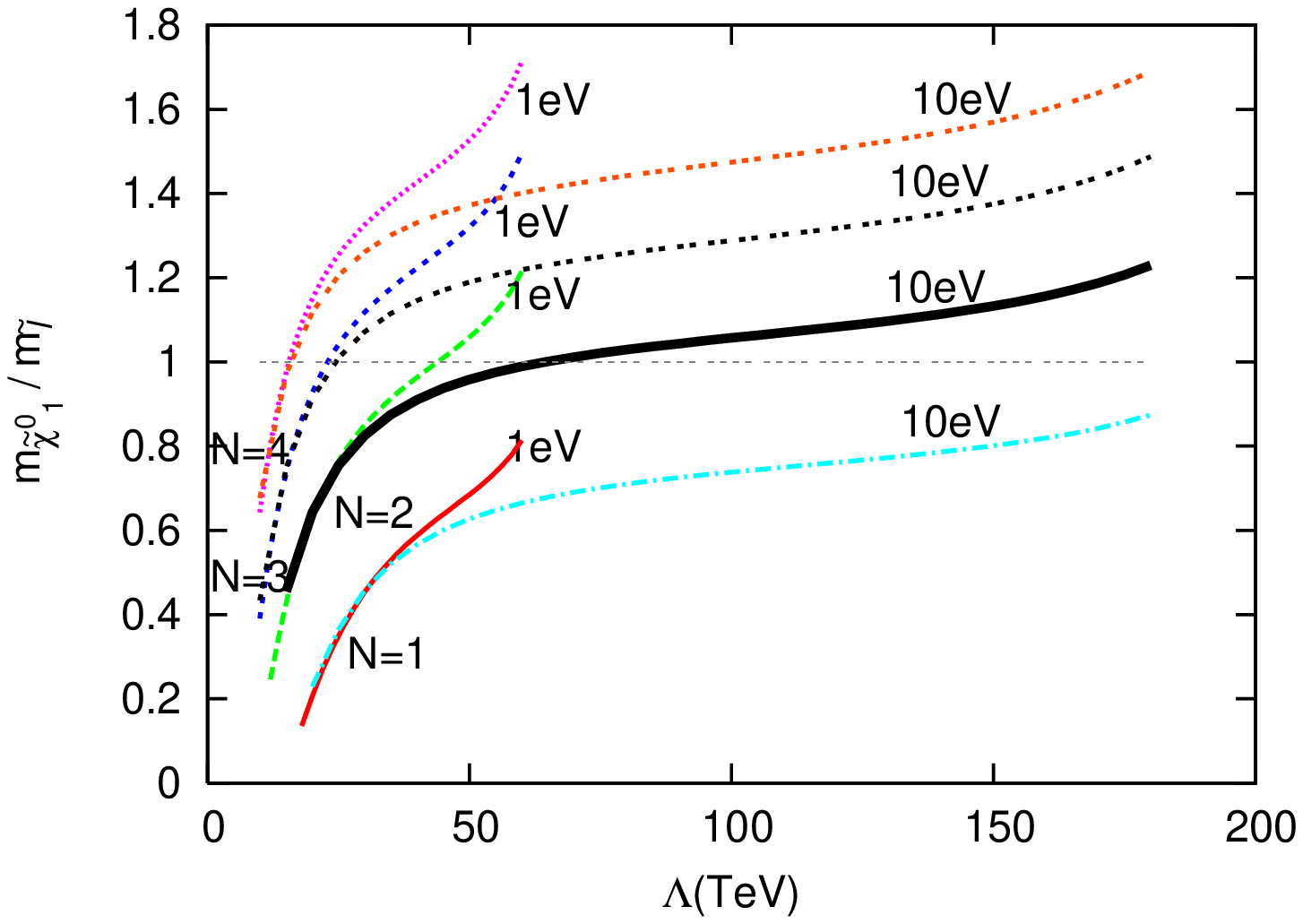,clip,width=5.5cm}}
\caption{$m_{\tilde{\chi}^0_1}/m_{\tilde{\ell}}$ dependence on $\Lambda$ for $\tan \beta=10$. These ratios are almost independent on $\tan\beta$}
\label{fig:mass_ratio}
\end{center}
\end{minipage}
\end{tabular}
\end{figure}
\begin{figure}[t]
\hspace{-3mm}
\begin{tabular}{cc}
\begin{minipage}{0.5\hsize}
\begin{center}
\rotatebox{-90}{\epsfig{file=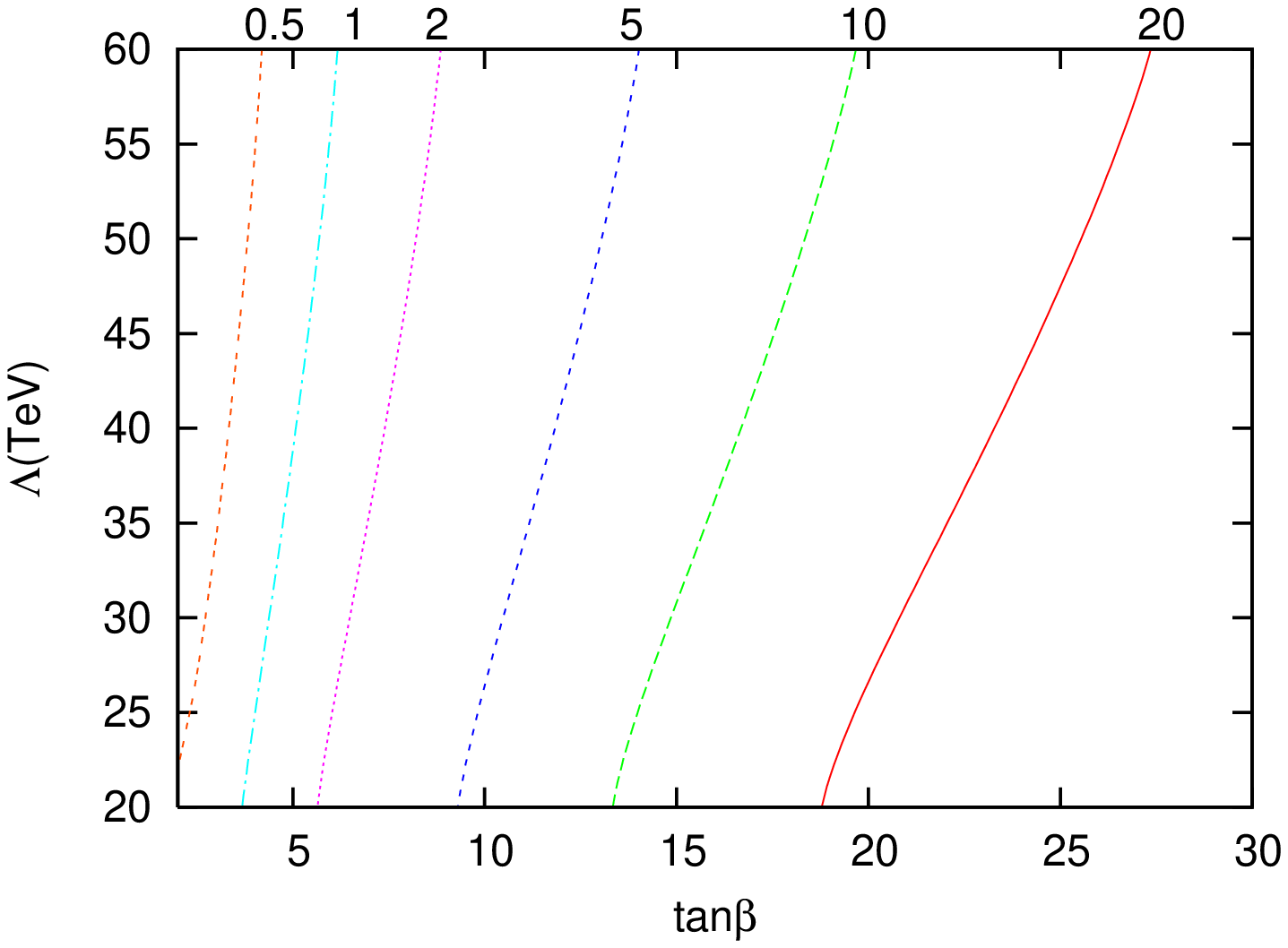,clip,width=5.5cm}}
(a)
\end{center}
\end{minipage}
\hspace{5mm}
\begin{minipage}{0.5\hsize}
\begin{center}
\rotatebox{-90}{\epsfig{file=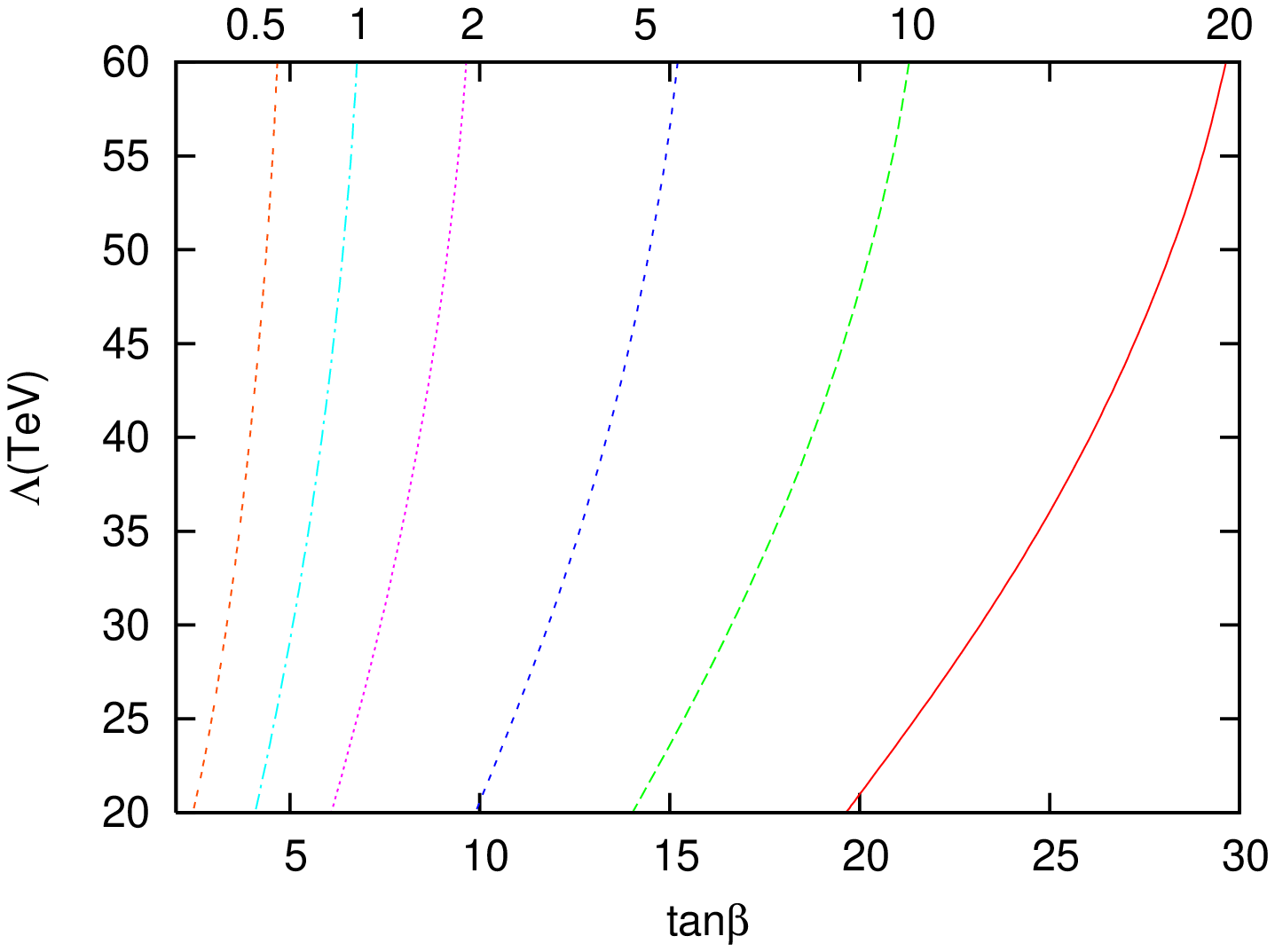,clip,width=5.5cm}}
(b)
\end{center}
\end{minipage}\\
\\
\begin{minipage}{0.5\hsize}
\begin{center}
\rotatebox{-90}{\epsfig{file=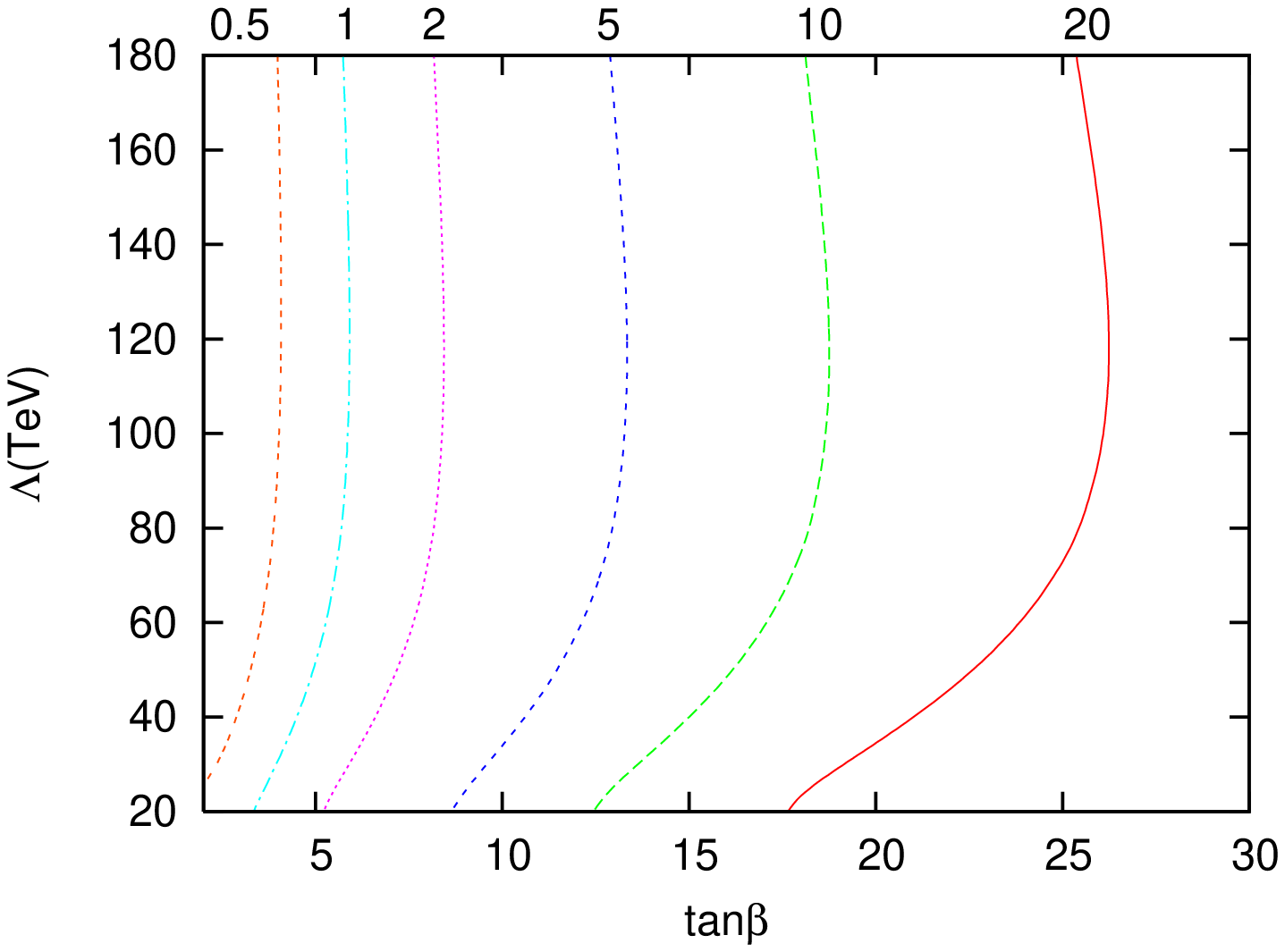,clip,width=5.5cm}}
(c)
\end{center}
\end{minipage}
\hspace{5mm}
\begin{minipage}{0.5\hsize}
\begin{center}
\rotatebox{-90}{\epsfig{file=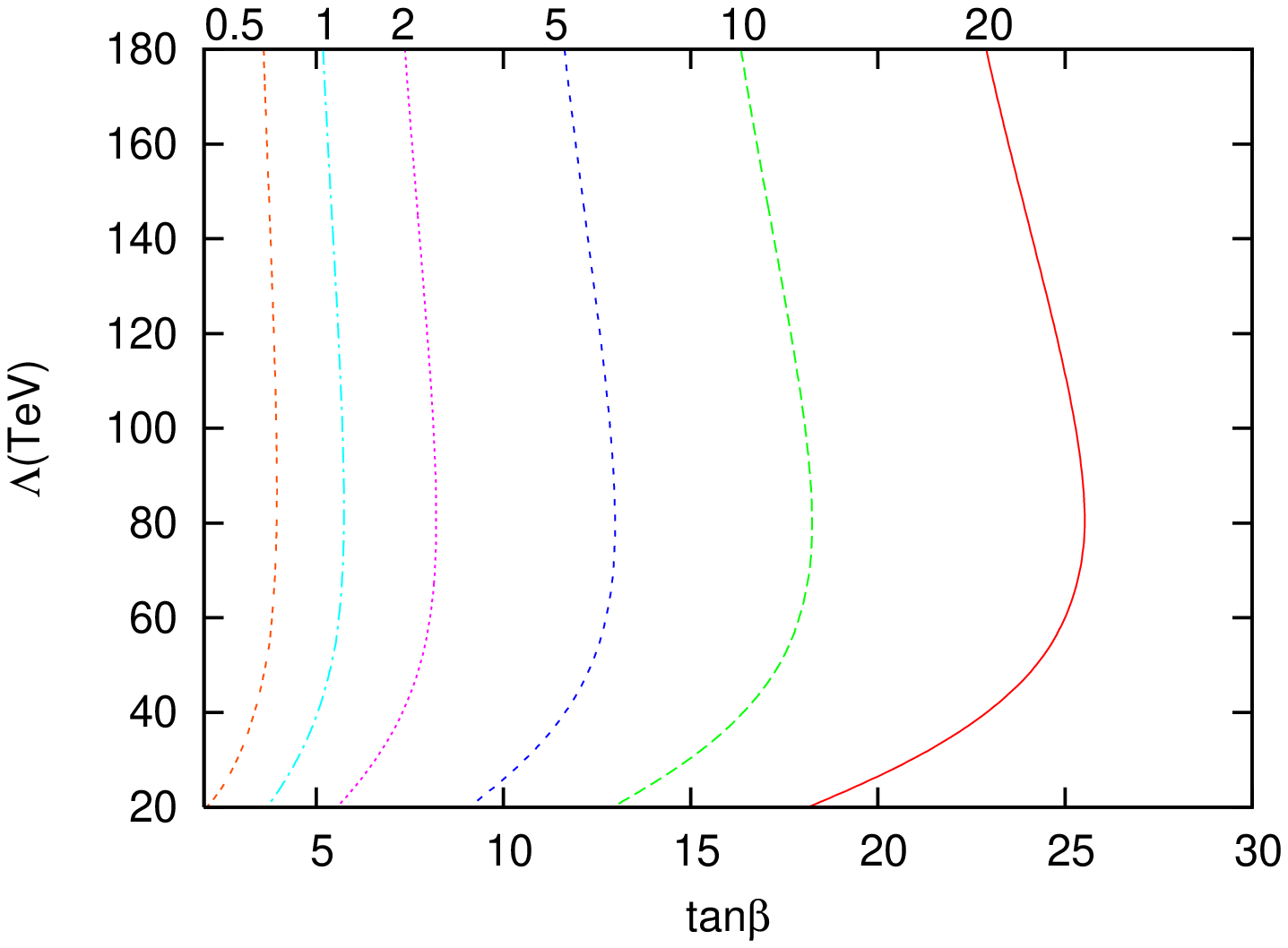,clip,width=5.5cm}}
(d)
\end{center}
\end{minipage}
\end{tabular}

\caption{The contour plots of the mass difference between $m_{\tilde{\tau}}$ and $m_{\tilde{\ell}}$,
$\Delta m\!=\!m_{\tilde{\ell}}-m_{\tilde{\tau}}$,
 on the ($\tan\beta$, $\Lambda$) plane. The contour parameters are $0.5$, $1$, $2$, $5$, $10$, $20$ GeV from left to right.
 (a) is for $N=2$ and $m_{3/2}=1$ eV, (b) $N=4$ and $m_{3/2}=1$ eV, (c) $N=2$ and $m_{3/2}=10$ eV, (d) $N=4$ and $m_{3/2}=10$ eV.
}
\label{fig:mass_diff}
\end{figure}

For our purpose, $m_{\tilde{\tau}}$, $m_{\tilde{\ell}}$ and $m_{\tilde{\chi}^0_1}$ are the most important parameters. 
In Fig.\ref{fig:slepton}, $m_{\tilde{\ell}}$ are shown as a function of $\Lambda$. We see that $m_{\tilde{\ell}}$ depend linearly on $\Lambda$. We have 
checked that $m_{\tilde{\ell}}$ are 
almost independent of $\tan\beta$.
Fig.\ref{fig:mass_ratio} shows the ratios
$m_{\tilde{\chi}^0_1}/m_{\tilde{\ell}}$. Those are also almost independent of $\tan\beta$, and change mainly through ${\cal O}(1)$ functions,
 $f(x)$ and $g(x)$ in Eq.(\ref{eq:gaugino_mass}) and Eq.(\ref{eq:scalar_mass}). 
For $N=1$, the mass spectrum cannot satisfy our assumption~($m_{\tilde{\tau}}<m_{\tilde{\ell}}<m_{\tilde{\chi}^0_1} $). 
Therefore we consider the case $N\ge 2$ in the present analysis.
Lastly, Fig.\ref{fig:mass_diff} shows the contour plots of $\Delta m=m_{\tilde{\ell}}-m_{\tilde{\tau}}$. 
We see that $\Delta m$ is a simply increasing function of $\tan\beta$. For $\Delta m < m_{\tau}$, the decay $\tilde{\ell}\rightarrow \ell+ \tau +\tilde{\tau}$ is
kinematically forbidden, and effectively all the $\tilde{\ell}$ decay  via $\tilde{\ell}\rightarrow \ell+ \tilde{G}_{3/2}$ for the ultra-light gravitino.
Thus, we concentrate on the case of $\Delta m>m_{\tau}$,
which corresponds to $\tan\beta \gsim 5$. 

 We should also note here on the constraints from the lower bound of the Higgs mass. In large parameter regions of our interest,
 the so-called decoupling limit is satisfied. Namely, CP-odd Higgs $A^0$ is heavy: $m_{A^0}\gsim200$ GeV.
Thus, the lighter CP-even neutral Higgs $h^0$ is an almost SM-like Higgs particle. 
We find that some parameter regions are excluded by the experimental lower bound $m_{h^0}\gsim 114$ GeV. 
For example, all parameter regions are excluded in the case of $N=2$ and $m_{3/2}=1$eV as shown in Fig.\ref{fig:gamma_ratio}-(a).

\begin{figure}[t]
\hspace{-3mm}
\begin{tabular}{cc}
\begin{minipage}{0.5\hsize}
\begin{center}
{\epsfig{file=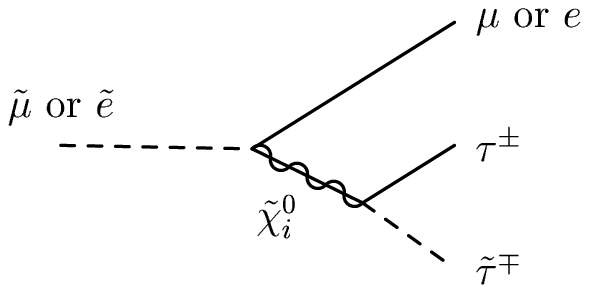, clip,width=5.5cm}}
\caption{A diagram for $\tilde{\ell} \rightarrow \ell + \tilde{\tau} + \tau$}
\label{fig:smudecay1}
\end{center}
\end{minipage}
\hspace{5mm}
\begin{minipage}{0.5\hsize}
\begin{center}
{\epsfig{file=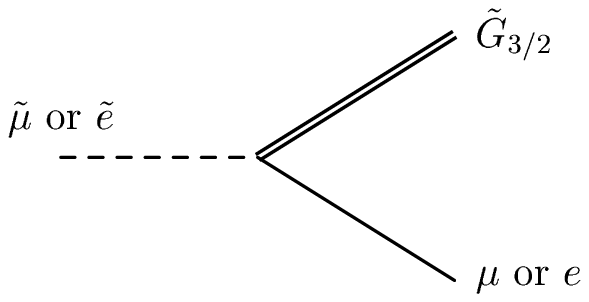,clip,width=5.5cm}}
\caption{A diagram for $\tilde{\ell} \rightarrow \ell + \tilde{G}_{3/2} $}
\label{fig:smudecay2}
\end{center}
\end{minipage}
\end{tabular}
\end{figure}

\section{$\tilde{\mu}/\tilde{e}$ decay}

In our model ${\tilde \mu}/{\tilde e}$ can decay in two channels, as shown in Fig.\ref{fig:smudecay1} and Fig.\ref{fig:smudecay2}. In the following, we neglect 
the masses of the electron and the muon.
The decay rate
$\Gamma(\tilde{\ell} \rightarrow \ell + \tilde{\tau} + \tau)$ is calculated from a Feynman diagram in Fig.\ref{fig:smudecay1}, where 
$\tilde{\ell} = \tilde{\mu}$ or $\tilde{e}$ and $\ell = \mu$ or $e$. 
Detailed calculations are given in \cite{Ambrosanio:1997bq}. For the parameter regions of our interest, 
the decay rate is approximately written as
\begin{eqnarray}
\Gamma(\tilde{\ell} \rightarrow \ell + \tilde{\tau} + \tau)&\approx&
\frac{2}{15\pi}\left(\frac{\alpha_{\rm EM}}{\cos^2\theta_{W}} \right)^2 m_{\tilde{\ell}}
\frac{(1+r_{\tilde{\chi}^0_1}^2)}{(r_{\tilde{\chi}^0_1}^2-1)^2}\left(\frac{\Delta  m}{m_{\tilde{\ell}}} \right)^5 \label{eq:3body1}\\
&=&4.4~\mathrm{eV} \left(\frac{m_{\tilde{\ell}}}{100~ {\rm GeV}} \right)^{-4} \label{eq:3body2}
\left(\frac{\Delta m}{10~ {\rm GeV}} \right)^5 \frac{1+r_{\tilde{\chi}^0_1}^2}{(r_{\tilde{\chi}^0_1}^2-1)^2},
\label{eq:Gamma3body}
\end{eqnarray}
where  $r_{\tilde{\chi}^0_1}=m_{\tilde{\chi}^0_1}/m_{\tilde{\ell}}$. Here we have assumed $\Delta m =m_{\tilde{\ell}}-m_{\tilde{\tau}}\gg m_{\tau}$. 
When $\Delta m$ becomes comparable to $m_\tau$, 
$\Gamma(\tilde{\ell} \rightarrow \ell + \tilde{\tau} + \tau)$
becomes even smaller than Eq.(\ref{eq:3body2}).
 On the other hand, 
$\Gamma(\tilde{\ell} \rightarrow \ell + \tilde{G}_{3/2} )$ is calculated from a diagram in Fig.\ref{fig:smudecay2}. 
For a light gravitino, 
the decay rate is given by
\begin{eqnarray}
\Gamma(\tilde{\ell} \rightarrow \ell + \tilde{G}_{3/2} )&=&\frac{1}{48\pi}\frac{m_{\tilde{\ell}}^5}{m_{3/2}^2M_P^2} \label{eq:G_decay}\\
&=&0.011~\mathrm{eV}~\left(\frac{m_{\tilde{\ell}}}{100 ~{\rm GeV}} \right)^5 \left(\frac{m_{3/2}}{1 ~{\rm eV}} \right)^{-2}.
\end{eqnarray}

\begin{figure}[t!]
\hspace{-3mm}
\begin{tabular}{cc}
\begin{minipage}{0.5\hsize}
\begin{center}
\rotatebox{-90}{\epsfig{file=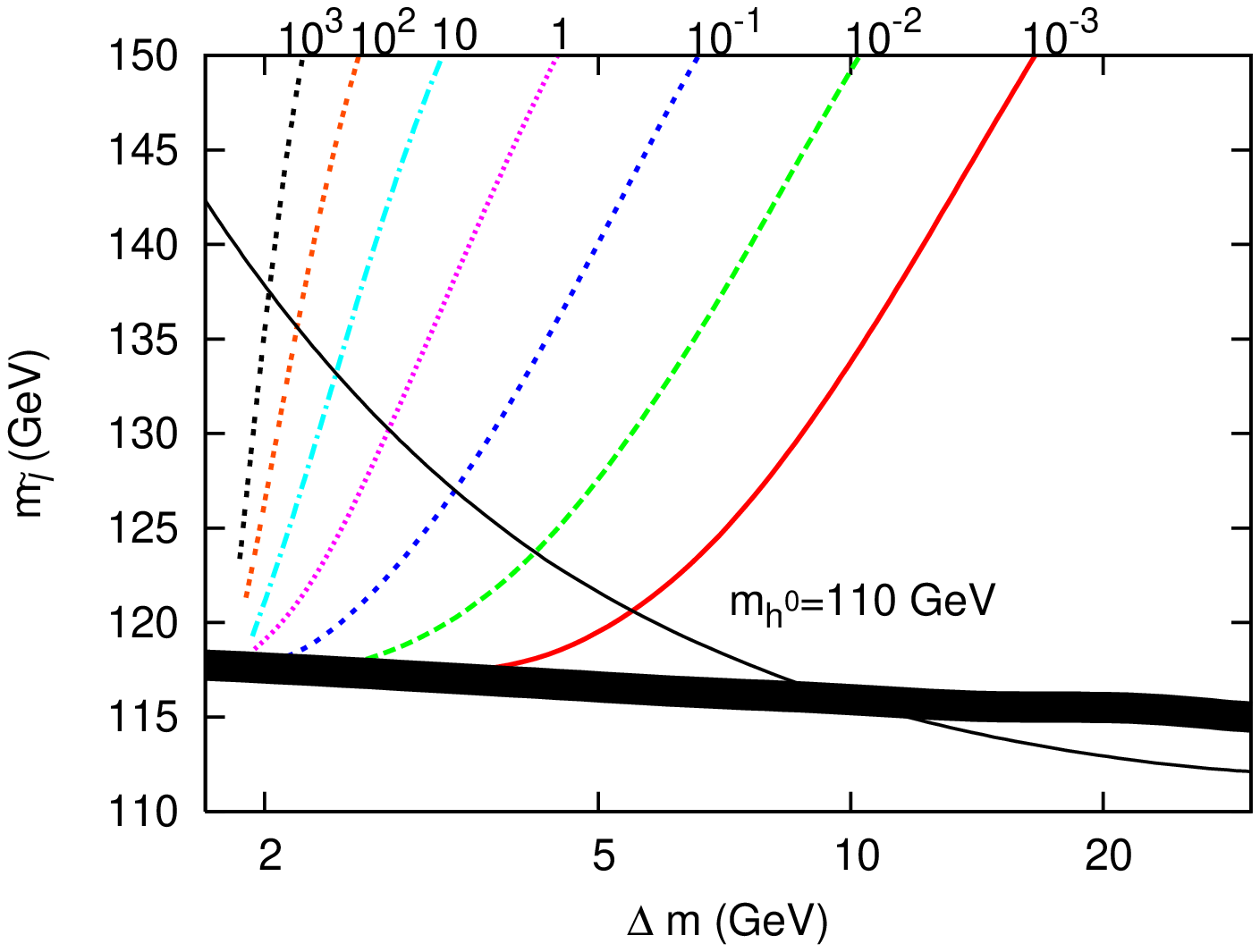,clip,width=5.5cm}}
(a)
\end{center}
\end{minipage}
\hspace{5mm}
\begin{minipage}{0.5\hsize}
\begin{center}
\rotatebox{-90}{\epsfig{file=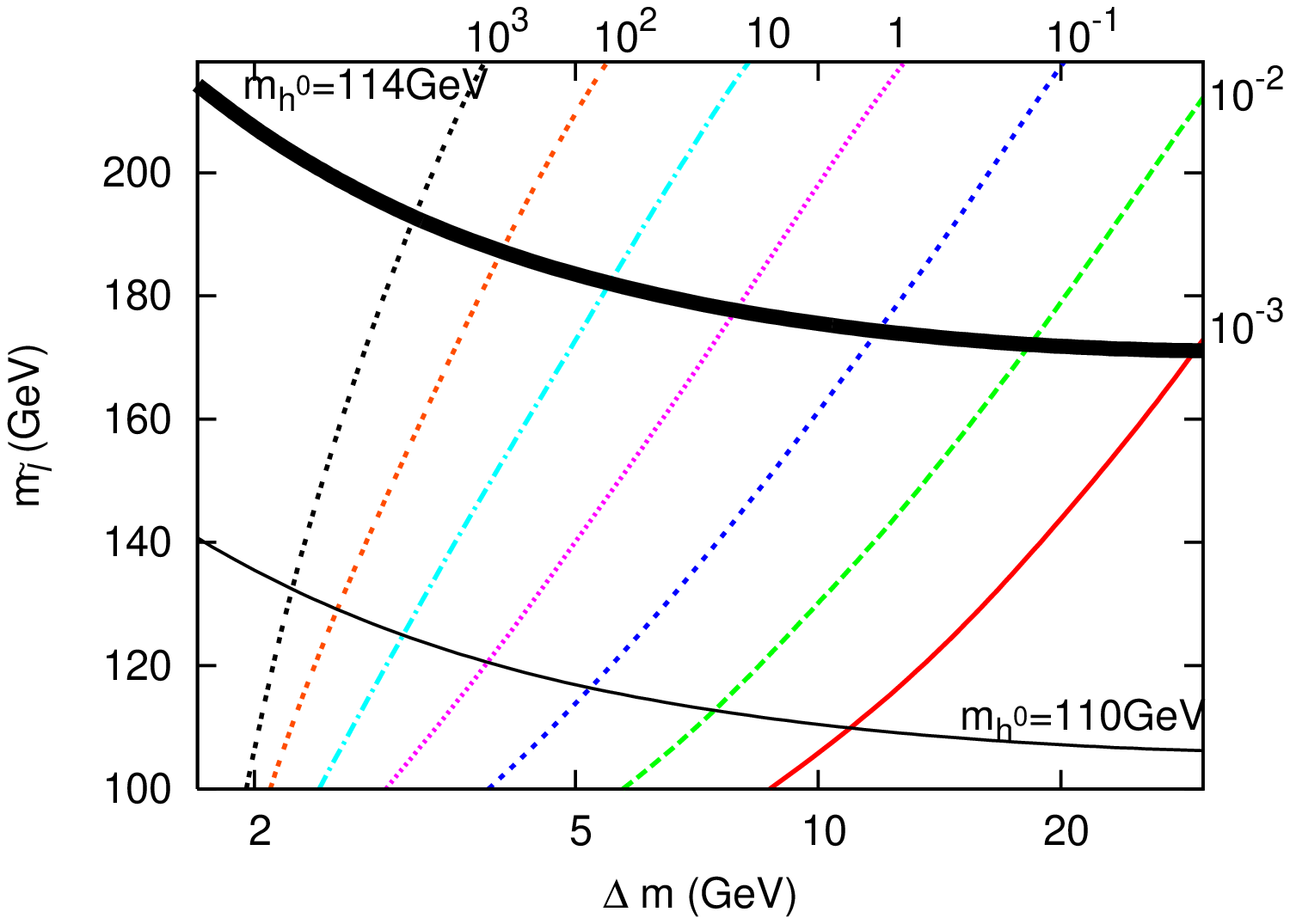,bb=251 55 554 487,clip,width=5.5cm}}
(b)
\end{center}
\end{minipage}\\
\\
\begin{minipage}{0.5\hsize}
\begin{center}
\rotatebox{-90}{\epsfig{file=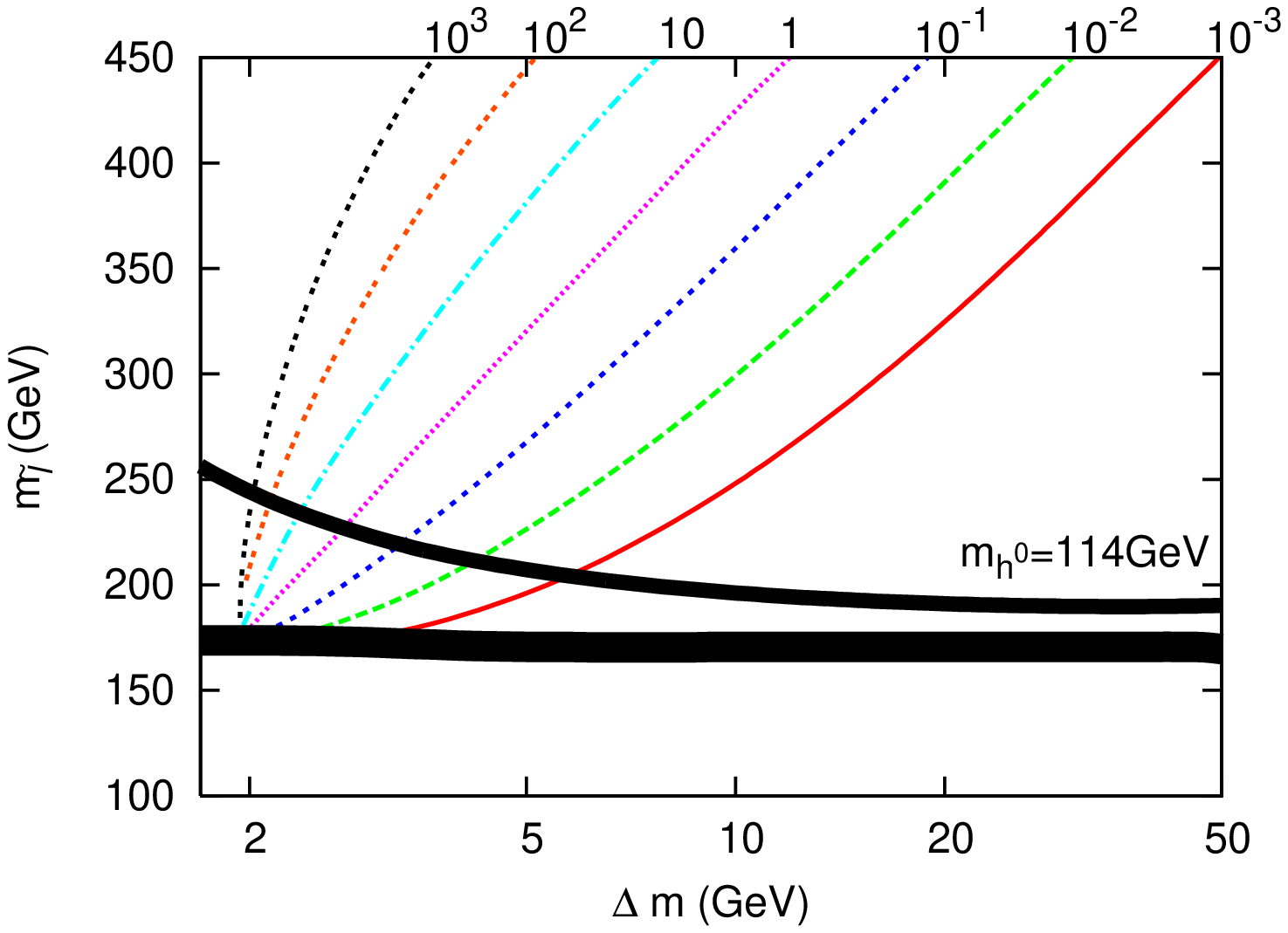,clip,width=5.5cm}}
(c)
\end{center}
\end{minipage}
\hspace{5mm}
\begin{minipage}{0.5\hsize}
\begin{center}
\rotatebox{-90}{\epsfig{file=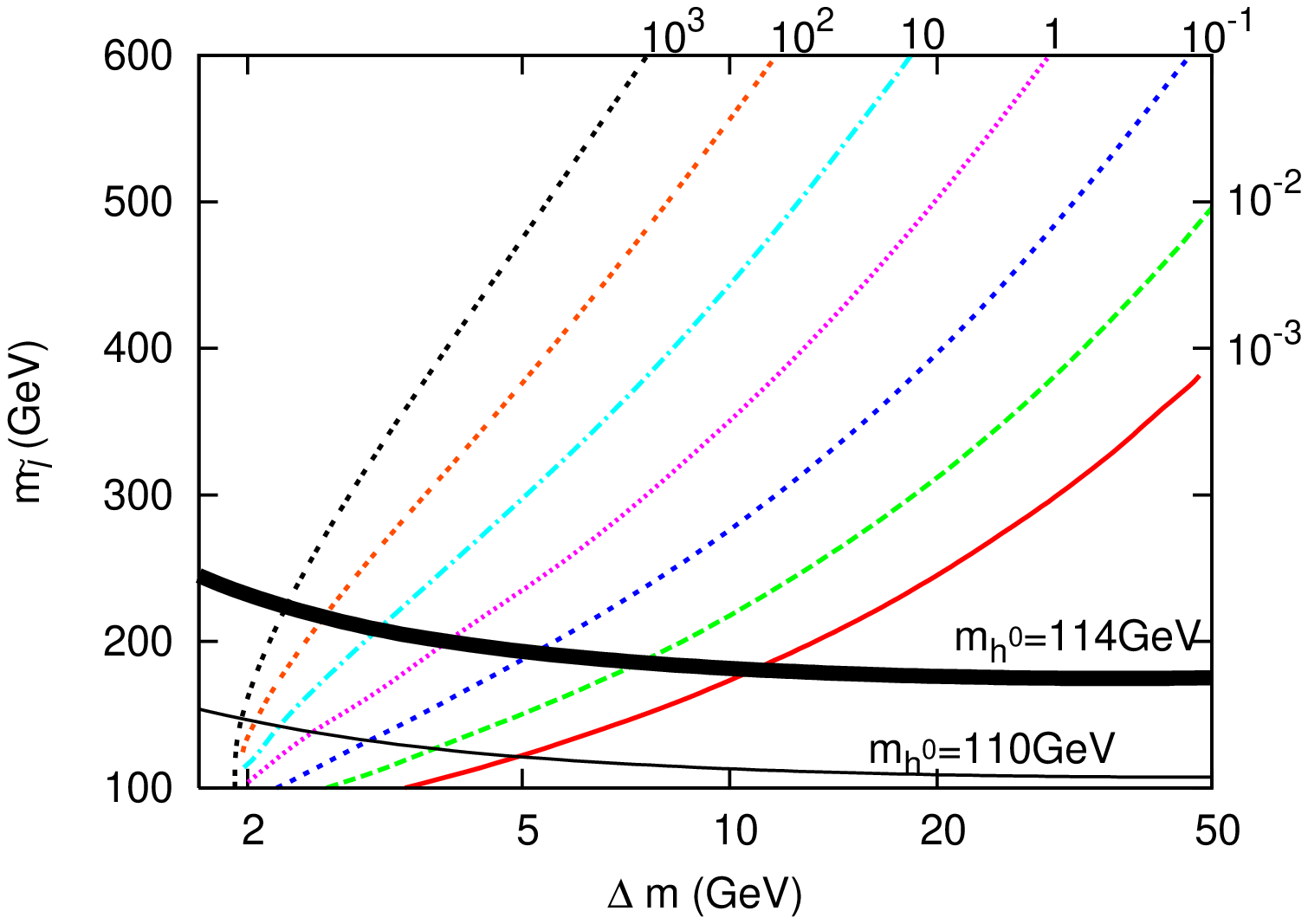,bb=251 55 554 487,clip,width=5.5cm}}
(d)
\end{center}
\end{minipage}
\end{tabular}
\caption{The contour plots of the ratio of the decay rates, 
$\Gamma(\tilde{\ell}\to\tilde{G}_{3/2}\ell)/\Gamma(\tilde{\ell}\to\tilde{\tau}\tau\ell) $,
 on the ($m_{\tilde{\ell}}$, $\Delta m$) plane. The contour parameters are $10^3$, $10^2$, $10$, $1$, $10^{-1}$, $10^{-2}$, $10^{-3}$. (a) is for $N=2$ and $m_{3/2}=1$ eV, (b)  $N=4$ and $m_{3/2}=1$ eV, (c) $N=2$ and $m_{3/2}=10$ eV, (d) $N=4$ and $m_{3/2}=10$ eV.
Below the thick solid line, the assumption $m_{\tilde{\chi}^0_1} > m_{\tilde{\ell}}$ is not satisfied. With the constraint of $m_{h^0}>114$GeV, all the region of (a) is
 excluded, and (b), (c) and (d) are partially excluded.}
\label{fig:gamma_ratio}
\end{figure}

We see that the decay rate $\Gamma(\tilde{\ell} \rightarrow \ell + \tilde{G}_{3/2} )$ becomes larger for smaller $m_{3/2}$. 
On the other hand, the decay rate
$\Gamma(\tilde{\ell} \rightarrow \ell + \tilde{\tau} + \tau)$ is suppressed for smaller $\Delta m$.
Therefore, in a certain parameter region, these two decay rates become comparable to each other. 
We have found parameter regions of $(N,\tan\beta,\Lambda)$ in which the ratios, 
$\Gamma(\tilde{\ell} \rightarrow \ell + \tilde{G}_{3/2} )/\Gamma(\tilde{\ell} \rightarrow \ell + \tilde{\tau} + \tau)$, 
are of ${\cal O}(1)$ (from $10^{-2}$ to $10^2$) as shown in Fig.\ref{fig:gamma_ratio}. From Eq.(\ref{eq:G_decay}), the gravitino mass is written as
\begin{equation}
m_{3/2}^2 = \frac{m_{\tilde{\ell}}^5}{48 \pi M_P^2}
\left(\frac{\Gamma(\tilde{\ell} \rightarrow \ell + \tilde{\tau} + \tau)}{\Gamma(\tilde{\ell} \rightarrow \ell + \tilde{G}_{3/2} )} \right)
\frac{1}{\Gamma(\tilde{\ell} \rightarrow \ell + \tilde{\tau} + \tau)}.\label{eq:G_mass}
\end{equation}
If masses for relevant SUSY particles are known, we can calculate the decay rate $\Gamma(\tilde{\ell} \rightarrow \ell + \tilde{\tau} + \tau)$ \cite{Ambrosanio:1997bq}.\footnote{For our parameter regions, 
$0.6 \lsim \Gamma(\tilde{\ell} \rightarrow \ell + \tilde{\tau} + \tau)/
\Gamma_{\rm no~mixing}(\tilde{\ell} \rightarrow \ell + \tilde{\tau} + \tau)\lsim 1 $,
where $\Gamma_{\rm no~mixing}$ is calculated without the mixing parameters. 
Thus, we can determine $m_{3/2}$ with an accuracy of factor 20\% 
without knowing the mixing parameters.}
Therefore, we can determine the gravitino mass $m_{3/2}$ from Eq.(\ref{eq:G_mass}) by combining the measured branching ratio, 
$\Gamma(\tilde{\ell} \rightarrow \ell + \tilde{G}_{3/2} )/\Gamma(\tilde{\ell} \rightarrow \ell + \tilde{\tau} + \tau)$, 
and the theoretical estimate of $\Gamma(\tilde{\ell} \rightarrow \ell + \tilde{\tau} + \tau)$.

\section{Summary}
Once SUSY particles are discovered at the LHC/ILC
and their masses are measured, one of the most important tasks in the next step would be 
the determination of SUSY breaking scale, or the gravitino mass.
In this letter we have proposed a new experimental method to determine the gravitino mass, which works well for an ultra-light gravitino of mass $m_{3/2}=1$--10~eV.
We have found that the branching ratio of $\Gamma(\tilde{\ell}\to\tilde{G}_{3/2}+\ell)/\Gamma(\tilde{\ell}\to\tilde{\tau}+\tau+\ell) $
 becomes $10^{-2}-10^2$ in a broad parameter
region for the gauge-mediated SUSY breaking scenario with the gravitino mass
$m_{3/2}=1$--10~eV.
We can estimate the gravitino mass with the measurement of this branching ratio, 
provided that the masses of the other relevant particles are known. 
The largest error of the gravitino mass may come from the 
error of $\Delta m$ in our method. To determine the gravitino mass
with an accuracy of factor 2,
 we need to know $\Delta m$ with an accuracy of factor 30\% (cf. Eq.(\ref{eq:Gamma3body})).
This may be accessible by precise measurements of the slepton masses and/or by studying the decay products of the process $\tilde{\ell}\to\tilde{\tau}+\tau+\ell$.

Remarkably, we have also found that,  the dominant decay modes
of the smuon and the selectron are $\tilde{\ell}\to \ell + \tilde{G}_{3/2}$
in a large parameter region. These processes
produce an energetic muon or electron together with missing energy, 
which may become clear signals for an ultra-light gravitino scenario.
A study of possible backgrounds in determining the ultra-light gravitino mass
at the LHC/ILC will be given elsewhere \cite{preparation}.

\end{document}